\documentclass[12pt]{article}
\def\e  {\epsilon}

\def\beq{\begin{equation}}
\def\eeq{\end{equation}}
\def\feq{\end{equation}}
\def\ee{\end{equation}}
\def\bea{\begin{eqnarray}}
\def\eea{\end{eqnarray}}
\def\bc{\begin{displaymath}}
\def\ec{\end{displaymath}}
\def\lb{\label}

\def\ds{ds^2=}

\def\th{\theta}

\def\lb{\label}

\def\ord#1{O\left(#1\right)}
\begin{document}
\title{Microscopic entropy of black holes and AdS$_2$ quantum 
gravity\footnote{\leftline{Talk given at the conference: {\it 
Black Holes in General
Relativity and String Theory}}
\leftline{\bigskip August 24-30 2008,
		 Veli Lo\v{s}inj,Croatia}}}

\author{Mariano Cadoni\footnote{E-mail: mariano.cadoni@ca.infn.it},
 Maurizio Melis\footnote{E-mail:maurizio.melis@ca.infn.it},
and  Paolo Pani\footnote{E-mail:paolo.pani@ca.infn.it}
\\
{\it  \small Dipartimento di Fisica, Universit\`a di Cagliari} \\
{\it\small and  I.N.F.N., Sezione di Cagliari,} \\
{\it\small  Cittadella Universitaria, 09042 Monserrato (Italy)}\\}

\vfill
\maketitle

\abstract{ Quantum gravity (QG) on two-dimensional anti-de Sitter
spacetime (AdS$_2$) takes always the form of a chiral conformal field
theory (CFT). However, the actual content of the CFT, and in
particular its central charge, depends on the background values of
the dilaton and Maxwell field. We review the main features of
AdS$_2$ QG with linear dilaton and of AdS$_2$ QG with
constant dilaton and Maxwell field. We also show that the 3D charged
Ba\~nados-Teitelboim-Zanelli black hole interpolates between these two
versions of AdS$_2$ QG. Applications to the computation of the
microscopic entropy of black holes are also discussed.}

\section{Introduction}
The problem of the microscopic origin of the Bekenstein-Hawking (BH)
entropy of black holes is one of the most intriguing  challenges for
modern theoretical physics.
Its solution is not only important for delivering  a  microscopic
basis for black hole thermodynamics. It also represents one crucial
test, perhaps the most relevant one, that any quantum  theory of
gravity has to pass.
It has been tackled using many different frameworks and  approaches:
String theory, AdS/CFT correspondence, asymptotic symmetries, D-branes,
induced gravity and entanglement entropy, loop quantum gravity.

Many of these approaches reproduce correctly the BH black hole
entropy (some exactly others  up to some numerical constant), in such
a good way that this success   is considered by some physicists
almost as a  problem \cite{Carlip:2007ph}.
It is likely that this universality, rather then a problem, is a
consequence of some fundamental underlying feature of semiclassical
quantum gravity that has to be shared by the different approaches. A
strong hint that this may be indeed the case is represented by the
wide, successfully, use of an asymptotical level formula for  
two-dimensional (2D) 
conformal field theory (CFT),
the so called Cardy formula, to count black hole microstates
\cite{Cardy:1986ie},
\beq\lb{cardy}
S=2\pi\left( \sqrt{c\,l_0\over 6}+\sqrt{c\,\bar l_0\over 6}\right),
\ee
where $l_{0}$ and $\bar  l_{0}$ are  the eigenvalues of the $L_{0}$ 
and $\bar L_{0}$ operators and  $c$ is the central charge in the   
conformal  algebra.

Obviously, Cardy's formula has a chance to reproduce BH black hole
entropy only if there is an underlying (at least approximate)
2D conformal symmetry.
This is for instance the case of black holes in  anti de Sitter (AdS)
spacetime.  The
AdS/CFT correspondence should allow us to describe black holes  as
thermal states of the dual CFT. 
An other approach is to use the built-in conformal symmetry of event
horizons and  2D diffeomorphisms  and the related 
algebra of
constraints, to model the black hole as a microstate gas of a CFT
( see e.g.  Ref. \cite{Carlip:1998wz}).

Counting microstates using the AdS/CFT correspondence  works well
only when the black hole  geometry factorizes as  AdS$_3 \times  M$
($M$ compact manifold) or AdS$_2\times  M$ .
For instance, the famous Strominger-Vafa \cite{Strominger:1996sh}
calculation of the entropy
of the 5D Reissner-Nordstr\"{o}m (RN) SUSY black hole has been made possible
because of the  AdS$_3$ factor in the near-horizon geometry of the 5D
black hole solution.
Precise computation of the statistical entropy of generic AdS$_d$
black hole (e.g. the Schwarzschild-AdS black hole in d=4) is out of
our reach, because  we do not know how to compute in strongly-coupled
gauge theories.

Thus, AdS$_3$ and AdS$_2$ QG together with the three and 
two-dimensional 
black holes  in  AdS  spacetime play a very special role for the
computation  of the statistical entropy of black holes. In the large $N$ limit
AdS$_{3}$ QG can be identified  as a 2D CFT
with central charge $c=3G/2l$ ($G$ and $l$ are respectively 
the 3D 
Newton constant and the AdS length)
describing Brown-Henneaux-like boundary excitations, i.e. deformations
of
the  asymptotic boundary of AdS$_{3}$ \cite{Brown:1986nw}.
The CFT  reproduces  correctly  the entropy of the Ba\~nados-Teitelboim-Zanelli 
(BTZ) black hole 
\cite{Strominger:1997eq} and of a wide class of higher-dimensional black holes.
 The related  thermodynamic system describes a 2D field theory with
extensive entropy  $S\sim T$  with a ground state of zero entropy at
zero
temperature.

On the other hand, it is still  not completely clear whether
AdS$_{2}$ QG  
has to be considered as the chiral half
of 2D CFT or a conformal quantum mechanics living on the asymptotic
one-dimensional  boundary of AdS$_{2}$ \cite{Cadoni:1998sg,Strominger:1998yg,
Maldacena:1998uz,Cadoni:1999ja,NavarroSalas:1999up,Cadoni:2000gm,
Hartman:2008dq,Sen:2008yk,Gupta:2008ki,Alishahiha:2008rt}. 
Nevertheless, it has been used with success  to compute the 
statistical entropy 
of AdS$_{2}$  black holes and related higher dimensional solutions.

One important application of  AdS$_{2}$ QG is its use in the 
description of  the 
near-horizon  regime of charged  extremal (BPS) black
holes, in which the  near-horizon geometry factorizes as 
 AdS$_2\times  M$.
In this case the dynamical system has peculiar features, such 
as the attractor mechanism \cite{Ferrara:1995ih,Ferrara:1996um,Sen:2005wa}, 
whereas from  the  
thermodynamical point of view the system is characterized by a ground
state of 
nonvanishing entropy at zero temperature.
Interesting examples of this kind of behavior are the near-horizon
geometries 
of asymptotically flat, extremal, black $p$-branes in $d$ space-time
dimensions,
\beq\lb{1}
AdS_{p+2}\times S^{d-p-2}=\frac{SO(p+1,2)}{SO(p+1,1)}\times 
\frac{SO(d-p-1)}{SO(d-p-2)}.
\feq
For $p=0$ we have charged, BPS, black holes in  $d=4,5$. 
For $p=1$ and  $d=5,6$ we have the  black string.
It is interesting to notice  that although Eq. (\ref{1}) holds also 
for $p=0$ and $d=3$,  this geometry cannot be obtained as the 
near-horizon geometry of a 3D charged black hole. This is because 
there are no asymptotically flat 3D black holes. 

The first, oldest, version of AdS$_2$ QG has been constructed 
following closely the Brown-Henneaux formulation of AdS$_3$ QG
\cite{Cadoni:1998sg}.
It is based on  AdS$_{2}$ endowed with a linear dilatonic background.
Recently, there  has been renewed interest for the AdS/CFT 
correspondence in two-spacetime dimensions 
\cite{Hartman:2008dq,Sen:2008yk,Gupta:2008ki,Cadoni:2008mw}.
In particular, a second formulation for  AdS$_2$ QG, based on  
AdS$_{2}$ endowed with constant dilaton  and Maxwell field has  been 
proposed in Ref. \cite{Hartman:2008dq}. 
In this paper we will argue  that the two different formulations of 
AdS$_{2}$ QG and their relationship with AdS$_{3}$
could be    the clue for understanding the  complicate pattern of 
near-horizon geometries of higher-dimensional charged black holes and 
their entropies. A key role in this context is played by
 3D charged
BTZ black hole. This  black hole interpolates 
between an asymptotic AdS$_{3}$  and a near-horizon 
AdS$_{2}\times S^{1}$ geometry. Circular symmetric dimensional 
reduction allows us to describe  AdS$_{3}$  as AdS$_{2}$ with a 
linear dilaton.  Thus, the charged BTZ black hole 
interpolates between the   two
different versions of AdS$_2$ QG. 

The plan of this paper is as follows.
In Sect 2 we give a  short review of the Brown-Henneaux formulation of 
AdS$_{3}$ quantum gravity.
In  Sect. 3 we briefly review AdS$_{2}$ QG with a linear dilaton.
In Sect. 4  we consider 
AdS$_{2}$ QG   with constant dilaton and
$U(1)$ field.
Some basic features of the charged BTZ black hole are discussed 
in Sect. 5.
In Sect. 6 we will show that charged BTZ black hole interpolates 
between the two formulations of AdS$_{2}$ QG.  
In Sect. 7 we discuss the application to the calculation of the 
microscopic black hole entropy.
Finally in Sect. 8 we state our conclusions.

\section{A  short review of AdS$_3$ quantum gravity}
Classical AdS$_{3}$ gravity is described by the action
\beq\lb{ac} I=\frac{1}{16 \pi G}\int d^{3}x
\sqrt{-g}\,( R+2\Lambda),\label{ac1} \ee
where $G$ is the 3D Newton constant
and $\Lambda=\frac{1}{l^2}>0$ is
the cosmological constant. We are using units where $G$
and $l$ have both the dimension of a length. 
Black hole solutions in AdS$_{3}$,   called BTZ after their 
discoverers  
Ba\~nados, Teitelboim and Zanelli \cite{Banados:1992wn,Banados:1992gq}, 
are characterized by mass 
$M$ and angular momentum $J$.
The corresponding line element in Schwarzschild coordinates is
\beq ds^2 =- f(r)dt^2
+ f^{-1}dr^{2}¥+r^2\left(d\theta -\frac{4GJ}{r^2}dt\right)^2,
\label{metric}\ee
with metric function: \beq
f(r)=-8GM+\frac{r^2}{l^2} +\frac{16 G^{2}¥J^2}{
r^2}.\label{metric2}
 \ee
The outer and inner horizons, $r_{+}$, $r_{-}$ 
 are given by 
 \beq
r^{2}_{\pm}={4Gl^2}\left(M\pm\sqrt{M^2 -
\displaystyle{\frac{J^2}{l^2}} }\right). \label{horizon1} \ee

AdS$_{3}$ quantum gravity was discovered by Brown and Henneaux 
\cite{Brown:1986nw} ten years
before Maldacena conjecture about the correspondence between 
gravity on AdS and conformal field theories \cite{Maldacena:1997re,Witten:1998qj}.  
They realized that the asymptotic symmetry group (ASG) of AdS$_{3}$,
i.e. the group that leaves invariant 
 the asymptotic form of the metric, is
the  conformal group in two spacetime dimensions. 

In order to determine the ASG one has first  to fix boundary
conditions for the fields at $r=\infty$ then to find the Killing
vectors leaving these boundary conditions invariant.
The boundary conditions  must be relaxed enough to allow for the
action of the conformal group and for the right boundary
deformations, but tight enough to keep finite the charges associated
with the ASG generators, which are given by boundary terms of the
action (\ref{ac}).
These charges can be
calculated using a canonical realization of the ASG
\cite{Brown:1986nw,NavarroSalas:1999sr}.
Alternatively, one can use a lagrangian formalism and work
out  the stress-energy tensor  for the boundary CFT
\cite{Balasubramanian:1999re}.  
For the  BTZ black hole suitable boundary conditions for the  
metric are
\cite{Brown:1986nw}
\bea\lb{bc}
g_{tt}&=& -\frac{r^{2}}{l^{2}}+\ord{1},\quad g_{t\theta}= \ord{1},\quad
g_{tr}=g_{r\th}= \ord{\frac{1}{r^{3}}},\nonumber\\
g_{rr}&=& \frac {l^{2}}{r^{2}}+\ord{\frac{1}{r^{4}}},\quad g_{\th\theta}=
r^{2}+\ord{1},
\eea
whereas the vector fields preserving them are
\bea\lb{vf}
&&\chi^{t}=l \left(\e^{+}(x^{+})+\e^{-}(x^{-})\right)+
\frac{l^{3}}{2r^{2}}(\partial^{2}_{+}\e^{+}+\partial^{2}_{-}\e^{-})+
\ord{\frac{1}{r^{4}}},\nonumber \\
&&\chi^{\theta}=\e^{+}(x^{+})-\e^{-}(x^{-})
-\frac{l^{2}}{2r^{2}}(\partial^{2}_{+}\e^{+}-\partial^{2}_{-}\e^{-})+
\ord{\frac{1}{r^{4}}},\nonumber \\
&&\chi^{r}=-r (\partial_{+}\e^{+}+\partial_{-}\e^{-})+
\ord{\frac{1}{r}},
\eea
where  $\e^{+}(x^{+})$ and $\e^{-}(x^{-})$ are arbitrary functions of
the light-cone coordinates $x^{\pm}= (t/l) \pm \th$ and
$\partial_{\pm}=\partial/\partial x^{\pm}$.
The generators $L_{n}$ ($\bar L_{n})$ of the diffeomorphisms with
$\e^{+}\neq 0$ ($\e^{-}\neq 0$) obey the Virasoro algebra,

\bea\label{va}
&&[L_{m},L_{n}]=(m-n) L_{m+n}+ \frac{c}{12}(m^{3}-m)\delta_{m+n\,0},\nonumber\\
&&[{\bar L}_{m},{\bar L}_{n}]=(m-n){\bar L}_{m+n}+\frac{c}{12}(m^{3}-m)
\delta_{m+n\,0},\nonumber\\
&&[L_{m}, \bar L_{n}]=0,
\eea
where $c$ is the central charge. In the semiclassical regime $c\gg 1$,
the  central charge can be calculated using a canonical
realization  of the ASG algebra.
Explicit 
computation of  $c$ gives
\cite{Brown:1986nw}
\beq\lb{cc}
c=\frac{3l}{2G}.
\ee
In a further development, Strominger  reproduced the entropy of the rotating, 
BTZ black hole  
 counting states of the Hilbert space of the CFT, i.e. using the Cardy
formula (\ref{cardy})
and identifying the eigenvalues of the  $L_{0}$ and $\bar L_{0}$ operators 
in terms of
the mass and angular momentum of the hole \cite{Strominger:1997eq},
\beq\lb{op}
lM= l_0+ \bar l_{0},\quad  J= l_0-\bar l_{0}.
\feq
Strominger calculation  holds for $c\gg 1$  and for large mass, large angular 
momentum black holes.
What  AdS$_{3}$ QG for $c\sim 1$  really  is, it is still not clear (see 
recent developments about this topic in Ref. \cite{Witten:2007kt,Maloney:2007ud})

\section{AdS$_2$ quantum gravity  with  linear dilaton}
The simplest theory of classical AdS$_{2}$ gravity contains a scalar field
(the dilaton $\eta$), para\-me\-tri\-zing (the inverse of)  2D Newton 
constant, 
\beq\lb{e1}
A=\frac{1}{2}\int\ d^2x\ \eta\left(R+\frac {2}{l^{2}}¥\right).    
\ee
The ensuing field equations do not allow for a constant dilaton but require a
linear dilaton background. Black hole solutions of mass $M$ are given 
by
\cite{Cadoni:1994uf}
\beq\lb{e2}
\ds-\left(\frac{r^2}{l^{2}}-\frac{2Ml}{\eta_{0}}\right)\,dt^2+
\left(\frac{r^2}{l^{2}}-\frac{2Ml}
{\eta_{0}}\right)^{-1}\,dr^2,
\qquad \eta=\eta_{0} \frac{ r}{l}.
\ee
The BH entropy of the 2D black hole is \cite{Cadoni:1994uf}
\beq\lb{bh}
S= 2\pi \eta_{h}=2\pi\sqrt{2\eta_{0}Ml},
\feq
where $\eta_{h}$ is the value of the dilaton at the black hole horizon.

Linear dilaton AdS$_{2}$ quantum gravity has been formulated following
closely
the Brown-Hen\-neaux derivation of AdS$_{3}$ QG
\cite{Cadoni:1998sg}.
 Suitable boundary condition for the metric and Killing vectors at the  
timelike boundary of AdS$_{2}$ are
\bea\lb{f2}
g_{tt}&=&-\frac{r^{2}¥}{l^2}+\ord{1},\quad g_{tr}=\ord{\frac{1}{ 
r^{3}}},\quad g_{rr}=
\frac{l^{2}¥}{r^2}+\ord{\frac{1}{r^4}},\\
\quad \eta&=&\ord{r},\quad \chi^t=\e(t)+\ord{\frac{1}{r^2}},
\quad\chi^r=- r\dot\e(t)+\ord{\frac{1}{r}}.
\eea
The ASG of AdS$_{2}$ is generated by one single copy of the Virasoro
algebra spanned by the  $L_{0}$ generators in Eq. (\ref{va}).
Thus AdS$_{2}$ quantum gravity can be seen as the chiral half of a 2D
CFT.
 The main difference between the AdS$_{2}$ and the AdS$_{3}$ case 
is the origin of the central charge $c$ in the Virasoro algebra 
(\ref{va}). In the 2D case the origin of the central charge can be traced back to 
the breaking 
of the $SL(2,R)$ isometry of AdS$_{2}$ owing to the linear dilaton 
background given by Eq. (\ref{e2}) \cite{Cadoni:2000ah}.

The  central charge  $c$ can be calculated using a canonical realization 
of the ASG algebra \cite{Cadoni:1998sg}\footnote{ The outcome of early 
calculations was  two times  the actual value of the central charge,
$c= 24\eta_{0}$ \cite{Cadoni:1998sg}. The origin of the  mismatch has been later
clarified in several independent ways
\cite{Cadoni:2000fq,Catelani:2000gn,Cadoni:2000gm}.},
 
\beq\lb{cc1}
c= 12\eta_{0}.
\ee
Using  Eq. (\ref{cc1}), identifying 
$l_{0}$ in terms of the black hole mass, $l_{0}=Ml$, 
the Cardy formula  (\ref{cardy}) reproduces  exactly 
the  entropy of the AdS$_{2}$ black hole given by Eq. (\ref{bh}). 

\section{AdS$_2$ quantum gravity with constant dilaton and U(1) field}

Recently Hartmann and Strominger  have found an
independent formulation of AdS$_{2}$ QG, which works for a background 
with  constant 
dilaton and differs in the mechanism generating the central charge 
\cite{Hartman:2008dq}.
The classical theory considered in Ref. \cite{Hartman:2008dq} is 2D 
Maxwell-Dilaton gravity,
\beq\lb{2da1}
 I=\frac{1}{2}\int d^{2}x \sqrt{- g¥}\left[\,
\eta \left(R+\frac{8}{l^{2}¥}¥\right)-
\frac{l^{2}}{4}  F^{2}\right], \ee
where $F_{\mu\nu}$ is the Maxwell tensor.
The ensuing equations of motion admit solutions describing AdS$_{2}$  
endowed with a constant dilaton and $U(1)$ field parametrized by a 
constant $E$. In the  conformal
gauge  the solutions  are given by
\beq\lb{h5}
ds^{2}= -\frac{l^{2}}{4\sigma^{2}¥}dx^{+}dx^{-},\, F_{+-}=2E 
{\mathbf\epsilon}_{+-},\,
A_{\pm}= \frac{El^{2}}{4\sigma},\, \eta=\frac{l^{4}E^{2}}{4},\,
\sigma=\frac{1}{2}(x^{+}-x^{-}).
\feq
We fix the diffeomorphisms and $U(1)$ gauge freedom  using a
conformal, 
respectively,  Lorentz gauge,
\beq\lb{gauge} ds^{2}=
-e^{2\rho}dx^{+}dx^{-},\quad \partial_{\mu}A^{\mu}=0. 
\feq 
After gauge fixing, conformal diffeomorphisms are described  by 
two arbitrary functions $\e^{+}(x^{+}¥),$ $\e^{-}(x^{-}¥$.
The  stress-energy tensor and the $U(1)$ current  are the 
constraints enforcing  gauge fixing  and generate, respectively, 
residual diffeomorphisms and gauge transformations
\bea\lb{h4}
&&T_{\pm\pm}=\frac{2}{\sqrt{- g}}\frac {\delta I}{\delta g^{\pm\pm}}=
-2\partial_{\pm}\eta\partial_{\pm}\rho + \partial_{\pm}\partial_{\pm}\eta
+2\partial_{\pm}h\partial_{\pm}a=0,\\
&& J_{\pm}=2\frac {\delta I}{\delta A^{\pm}}=\pm 2\partial_{\pm} h=0,
\eea 
where $h$ is an auxiliary field used to linearize  the quadratic term 
for the $U(1)$ field and we have dualized the vector potential 
$A_{\mu}$ in 
terms of a scalar $a$.
If one now requires that the asymptotic boundary of AdS$_{2}$ 
remains at $\sigma=0$ under the action of conformal diffemorphisms 
(this is equivalent to fix boundary conditions for the metric)
$\e^{-}$ is 
determined in terms  of $\e^{+}$. We are left  with only a chiral half of 
the 2D  CFT.
Analogously to the previous realization of AdS$_{2}$ QG the symmetry
algebra is 
one single copy of the Virasoro algebra.

Being  the dilaton constant, one is led to conclude that we are 
dealing with pure 2D QG, which has vanishing  central charge 
\cite{Distler:1988jt}.
This is not the case because of the presence of the $U(1)$ field.
We need boundary conditions for the vector
potential  at $\sigma=0$. 
Absence of charged current flow out of the boundary of AdS$_{2}$ requires
$A_{\mu}(\sigma=0)¥=0$. 
The problem is that this boundary condition is not invariant under the action of
conformal diffeomorphisms,
\beq\lb{h7}
\delta_{\e}A_{\mu}|_{\sigma=0}=\frac{l^{2}E}{2}\partial^{2}¥_{+}\e^{+}
|_{\sigma=0}. \feq
This term can be cancelled by a gauge transformation $A\to A 
+\partial \lambda$ with
 \beq\lb{h71}
\lambda(x^{+})=-\frac{l^{2}E}{2}\partial_{+}\e^{+}.
\feq
Hence, the conformal symmetry group is  a chiral half  of conformal 
diffeomorphisms supplemented by the gauge transformation (\ref{h71}). 
We have a twisted CFT. Conformal transformations are generated by 
Virasoro generators given in terms of an improved  stress energy 
tensor,
\beq\lb{iset}
\tilde L=\frac{1}{2}\int dx^{+}\tilde T_{++}\e^{+}, \quad 
\tilde T_{++}=  T_{++}+ \frac{E l^{2}}{4}\partial_{+}J^{+}.
\feq
The central charge in the Virasoro algebra can be calculated
expanding in  Laurent modes and using the  transformation law of the 
improved stress-energy tensor
\beq\lb{h16}
\delta_{\e}\tilde T_{--}= \e^{+}\partial_{+}\tilde 
T_{++}+2\partial_{+}
\e^{+} \tilde T_{++}+\frac{c}{12}\partial_{+}^{3}\e^{+}.
\feq
The transformation law of the original $T_{++}$ is anomaly-free, but
that of the current $J_{+}$ may have an anomalous term proportional to
its level $k$ \cite{Hartman:2008dq},
\beq\lb{h18}
\delta_{\lambda}J_{+}= k\partial_{+}\lambda^{+}.
\feq
This produces a central charge $c$ in  the Virasoro
algebra given by,
\beq\lb{h20}
c=3k E^{2}l^{4}=\frac{3}{4} k\sqrt{\frac{\pi}{G}}\, l Q.
\feq
 Let us close this sections by summarizing the main results of the last 
 two sections. We have  two different 
 formulations of  AdS$_{2}$ QG; both 
 are described by the chiral half of a 2D CFT but the origin of the 
 central charge is drastically different. In the first case,  AdS$_{2}$ with a 
 linear dilaton, the central charge is originated by the breaking of 
 the conformal symmetry caused by a nonconstant dilaton and is
 determined by 2D inverse Newton constant 
 $\eta_{0}$. 
 In the second case, AdS$_{2}$ with a 
 constant dilaton and $U(1)$ field,   the central charge is produced 
 by a Schwinger effect and by a twisting of the CFT and  is determined by 
 the electric field $E$. 
 To find a bridge between  the two  formulations we have to go up to 
 three dimensions and to bring into  the play the charged BTZ black hole.

\section{The charged BTZ black hole}
AdS gravity in three spacetime dimensions admits also  charged black 
hole solutions,  
which are the charged version of the BTZ black hole \cite{Martinez:1999qi}. 
They are solution of the
action
\beq\lb{action1}
I=\frac{1}{16\pi G}\int d^{3}x \sqrt{-g}\,
(R+\frac{2}{l^{2}}-4\pi G F_{\mu\nu}F^{\mu\nu}) \label{ac2},
\ee
where $F_{\mu\nu}$ is
the electromagnetic (EM)  field strength.
Considering for simplicity solutions with zero angular momentum, we
have the two-parameter  $(M,Q)$ family of electric charged  black hole  solutions 
\cite{Martinez:1999qi}
\bea\label{metric3}
  ds^2 &=&- f(r)dt^2
+ f^{-1}dr^{2}¥+r^2d\theta^2,\nonumber\\ 
f(r)&=&-8GM+\frac{r^2}{l^2}
-8\pi GQ^2 \ln (\frac{r}{w}),\quad  F_{tr}=\frac{Q}{r},
\eea
where $M,w$ are constants 
and $-\infty<t<+\infty$, $0\leq r<+\infty$,
 $0\leq \theta \leq2\pi$. Notice that the parameter $w$ can be 
 reabsorbed in the definition of $M$. 
The striking differences with the BTZ black hole  is represented by the 
presence of  a power-law
curvature singularity at $r=0$.
The charged BTZ black hole has an inner, $r=r_{-}$, and outer, 
$r=r_{+}$,  event horizon.
It also has   well-defined  temperature and entropy,
\beq \label{tem}T_H=\frac{r_+}{2\pi l^2}-\frac{2 G
Q^2}{r_+}, \quad S=\frac{\pi r_+}{2G}= \frac{\pi l}{G} \sqrt{2GM+ 2\pi
GQ^{2}\ln(\frac{r_{+}}{w}})
.\ee

The charged BTZ black hole has been considered as the Cinderella in
the family of 3D AdS black hole celebrities. The reason is that  it
has some unpleasant  features.
By varying the action one gets logarithmic divergent boundary terms. This
makes the mass of the solution a poorly defined  concept.
Moreover, in order to avoid naked singularities one must impose 
a BPS-like bound  involving $M$ and $Q$, 

\beq\lb{delta}
\label{ev}\Delta=8GM-4\pi G Q^2[1-2\ln(\frac {2Ql}{w}\sqrt{\pi
G})] \ge 0.\ee
Unfortunately, this bound can be satisfied for arbitrary negative values of $M$,
making the definition of thermodynamic  ensembles  problematic.

These problems  can be handled using
renormalization group ideas and the IR/UV relation for the  
AdS/CFT correspondence \cite{Cadoni:2008mw,Cadoni:2007ck}. 
The system is enclosed in a circle of radius $r_{0}$ (the UV cutoff for
the CFT), one takes $r,r_{0}¥\to \infty$,
keeping the ratio $r/r_{0}=1$, and writes, 
\beq\label{mass}
f(r)=-8GM_{0}(r,w)¥+\frac{r^2}{l^2}
-8\pi GQ^2 \ln (\frac{r}{r_{0}¥}),\quad M_{0}(r_{0},w)=M+\pi Q^{2}¥\ln(\frac{r_{0}}{w}).
\ee
The parameter $w$ is interpreted as a running scale
and $M(r_{0}¥,w)$ is the regularized black hole mass, the total 
energy (gravitational plus 
electromagnetic)  inside a circle of radius $r_{0}$.
Basically, one has now two options:
\begin{enumerate}
\item
$M$ is kept fixed and the metric (hence the position of the horizon)
is scale-dependent. In this case $M$ is seen as the black hole  mass  \cite {Martinez:1999qi}.
\item 
The metric (hence the horizon position) is $w$-independent   
and $M$ runs with $w$ \cite{Cadoni:2008mw,Cadoni:2007ck}.
\end{enumerate}
Because we want to keep the horizon (the IR scale for the CFT) fixed, we use 
prescription $2$. $M$ runs with $w$: $w\to \lambda w,\,
M\to M+\pi Q^{2}\ln\lambda,$
but $M_{0}$  is 
$w$-independent. We  fix now $w=l$ and 
$r_{0}¥=r_{+}$.  
The invariant black hole mass,  to be identified with the conserved 
charge associated with time-translations, becomes
\beq\label{mass1}
M_{0}(r_{+},l)=M+\pi Q^{2}¥\ln(\frac{r_{+}}{l}).
\ee
This solves the problem of divergent boundary terms in the variation 
of the action (\ref{action1}). Moreover, the use of the mass $M_{0}$ 
of Eq. (\ref{mass1}) instead of $M$ as the energy of the system 
allows for a consistent formulation 
of the thermodynamics  of the charged BTZ black hole  \cite{cristina}. 

\subsection{The near-horizon limit}
We expect the generic near-horizon, extremal behavior of black
branes given by Eq. (\ref{1})  
to hold also for $p=0$ and $d=3$   not for asymptotically flat  but for 
asymptotically AdS black 
holes.
Thus, we expect an AdS$_{2}\times S^{1}$ near horizon geometry 
for our extremal charged BTZ black hole.

In the extremal limit the charged BTZ black hole saturates the bound 
(\ref{delta}), i.e. we have  $\Delta=0,\, r_{-}=r_{+}=\gamma= 2\sqrt{\pi G} 
Q l$.
Expanding near the horizon, $r= \gamma+x$ one finds that the 3D geometry
factorize 
as AdS$_{2}¥\times S^{1}$, whereas the  EM field becomes constant,
\beq\lb{17a}
ds^{2}¥=-fdt^{2}+
f^{-1}¥dx^{2}+ 
\gamma^{2}d\theta^{2}¥,\, f=\left(\frac{2}{l^{2}} x^{2}- 8G\Delta 
M \right),\,
F_{tx}=
\frac{1}{2\sqrt{\pi G}\,l},
\feq
where $\Delta M= M-M(\gamma)= M- \pi Q^{2}( \frac{1}{2}-
\ln(2Q\sqrt{\pi G}))$ is the  mass above extremality.
This black hole solution shares with its higher-dimensional, 
asymptotically flat, cousins the
thermodynamical
behavior. The extremal charged BTZ black hole  is a state of zero
temperature and 
constant entropy  $S_{ext}= \pi\gamma/2G$. 
Thus, the charged BTZ black hole  interpolates between an asymptotic, 
$r\to\infty$, AdS$_{3}$ geometry
and a near horizon AdS$_{2}¥\times S^{1}$ geometry.

\section{Interpolating the two versions of AdS$_2$ quantum gravity}

The two limiting regimes, the asymptotic and near-horizon one, 
of the BTZ black hole can be both described by an effective 2D
Maxwell-Dilaton theory of gravity. 
The 2D effective theory can be obtained performing a circular 
symmetric dimensional reduction 3D$\to$ 2D, with the dilaton 
parametrizing the radius of the transverse circle and 
with an  electric  ansatz for the Maxwell field,
$F_{t\theta}=F_{r\theta}=0$,
\beq\lb{dr}
ds_{(3)}^{2}= ds_{(2)}^{2}+l^{2}\eta^{2}d\theta^{2}.
\feq
The 2D Maxwell-Dilaton  gravity  theory  turns out to be,
\beq\lb{2da}
 I=\frac{1}{2}\int d^{2}x \sqrt{-g¥}\,
\eta \left(R+\frac{2}{l^{2}}-4\pi G F^{2}\right) . \ee
The corresponding 2D field equations admit two classes of solutions whose metric
part is always a 2D AdS spacetime: 
\begin{itemize}
\item
 AdS$_{2}$ with linear dilaton and Maxwell field $F_{tr}=Q/r$.
This corresponds to the asymptotic $r\to\infty$ regime of the charged 
BTZ black hole.
\item
 AdS$_{2}$ with constant dilaton and electric field. This corresponds to the
near horizon regime.
\end{itemize}

\subsection{AdS$_{2}$ with linear dilaton}
These solutions  are nothing but the  3D solution written in a 2D form. 
They are  given by
the 2D sections  of the 3D solutions (\ref{metric3}) and with 
$\eta=\bar\eta_{0}(r/l)$. 
Owing to a scale symmetry, $\eta\to\lambda \eta$, of the 2D field 
equations, the
constant $\bar\eta_{0}$  is determined by the dimensional reduction:
\beq\lb{k4}
\bar\eta_{0}=\frac{l}{4G}.
\feq
Mass, temperature and entropy of the 2D black hole  are  the same as  those
of the 3D black hole.
 
The dual CFT can be constructed following the same procedure used in 
Sect. 3
for  2D dilaton gravity without Maxwell  field. 
There is, however, a non trivial detail.  Not
only the charge associated with the $L_{0}$ Virasoro operator (the mass)  
 diverges, but also the
other charges associated with the  other Virasoro operators $L_{m}$. 
The renormalization procedure used in the previous section
for the mass allows also to
define finite 
charges for the other Virasoro operators (see for details 
Ref. \cite{Cadoni:2007ck}). 
It turns out that the central charge 
 of the Virasoro algebra  is also finite and matches exactly that of 
 pure AdS$_{2}$ with linear dilaton, 
 \beq\lb{k8}
 c=12\bar\eta_{0}=\frac{3l}{G}.
 \ee
The EM field does not contribute to the central charge but  only
enters in the renormalization of the eigenvalue of $L_{0}$, which is 
given in terms the mass $M_{0}$ of Eq. (\ref{mass1}),
\beq\lb{k10}
L_{0}=lM_{0}(r_{+},l).
\ee

\subsection{AdS$_{2}$ with constant dilaton and electric field}
The 2D field equations stemming from the action (\ref{2da})
admit also solutions describing AdS$_{2}$ with
constant dilaton and electric field. They are 
given  by the 2D
sections  of the near-horizon 3D solution (\ref{17a}) . 
 A Weyl transformation of the metric together with a  rescaling by a
constant of the  $U(1)$ field strength brings the 2D action into the
form \cite{Cadoni:2008mw},
\beq\lb{2da2}
 I=\frac{1}{2}\int d^{2}x \sqrt{- g¥}\left[\,
\eta \left(R+\frac{(\partial\eta)^{2}}{\eta}+
\frac{2\eta}{l^{2}\eta_{0}¥}\right)-
\frac{l^{2}}{2}  F^{2}\right]. \ee 
The classical  solutions are
\bea\lb{nh}
ds^{2}&=& -( \frac{2}{l^{2}}x^{2}-a^{2}) dt^{2}+ (
\frac{2}{l^{2}}x^{2}-a^{2})^{-1}¥ dx^{2}, \quad
 F_{\mu\nu}=2E \varepsilon_{\mu\nu},\nonumber\\
\quad \eta&=&2l^{4}E^{2},\quad\quad  E^{2}= \frac{1}{4 l^{3}} 
\sqrt{\frac{\pi}{G}}\, Q.
\eea

Apart from a trivial redefinition of the AdS length, this 2D model
differs from the Hartmann-Strominger model just for the presence of a
kinetic term  for the dilaton and a dilaton potential $V(\eta)$.
In a constant dilaton background these terms do not contribute to
the central charge.
 It is a simple exercise to construct the dual twisted CFT describing
AdS$_{2}$ QG using the Hartmann-Strominger procedure  described in 
Sect. 4 (see for details Ref.
\cite{Cadoni:2008mw}).
The central charge of the twisted CFT turns out to be
\beq\lb{h21}
c=3k E^{2}l^{4}=\frac{3}{4} k\sqrt{\frac{\pi}{G}}\, l Q.
\feq

\section{Microscopic black hole entropy}

We can easily reproduce the Bekenstein-Hawking entropy of the 2D AdS
black hole and hence the entropy of the charged BTZ black hole 
calculating the asymptotic density of states for the linear dilaton
CFT. Using Eqs. (\ref{k8}), (\ref{k10}) and (\ref{mass1}) into the 
Cardy formula (\ref{cardy}) we find exactly the BH entropy (\ref{tem}).

In principle, one should also be able to reproduce the entropy of the extremal
(and near-extremal) charged BTZ black hole  by calculating
the asymptotic density of states for the twisted  CFT.
 However  this requires using in the Cardy formula the eigenvalues of
the twisted operator $\tilde L_{0}$ instead of that for the untwisted one.
Careful analysis of the CFT spectrum and detailed knowledge of the
effect of twisting on the CFT Hilbert space is needed. 
\section{Conclusions}

The two different realizations of AdS$_{2}$ QG investigated in this 
paper describe  different states:
AdS$_{2}$ QG with linear dilaton  describes  Brown-Henneaux-like boundary 
excitations, which
are relevant for explaining the entropy of the BTZ black hole whereas 
AdS$_{2}$ QG
with constant dilaton and maxwell field describes D-brane-like 
excitations,
which should account correctly
for the entropy of extremal BPS black holes. 
  Both realizations have a dual  gravitational description in terms
of an asymptotic and near-horizon geometry.
  Similarly to what happens for higher-dimensional  charged RN
solutions, there is an interpolating gravitational solution, the
charged BTZ black hole bridging the two descriptions.
 These features make AdS$_{2}$ QG a powerful tool for investigating
microscopic black hole physics and to shed light  on several features
of the AdS/CFT correspondence.

There is a long list of open questions and possible further developments.
One should be able to reproduce the entropy of extremal and near-extremal  (BPS) 
black holes using the near-horizon CFT. 
From the gravitational side  this requires the use of the entropy 
function formalism \cite{Sen:2008yk,Alishahiha:2008tv}, whereas from the CFT side 
requires careful investigation 
of the Hilbert space of the twisted CFT.

An other key issue  is the understanding, at the pure 2D level, of 
the relationship between the 
two sectors of 2D Maxwell-Dilaton gravity, the one with constant 
dilaton and the other with linear varying dilaton. In Ref. \cite{Castro:2008ms} 
it has been shown that the constant dilaton sector requires  a 
negative 2D Newton constant.  A true unified description of both constant and 
linear dilaton sector would shed light on these issues.

Also from the CFT point of view the relationship between the 
asymptotic CFT and the near-horizon CFT is  far from being understood.
The relevant question here is whether or not these two realizations 
correspond to two different conformal points.
In the case of 3D AdS gravity minimally coupled with 
a scalar field it has been shown that the two dual CFTs are related 
by renormalization group flow and that the Zamolodchikov $c$-theorem holds 
\cite{Hotta:2008xt}. Presently it is not clear if the same holds for 
Maxwell-Dilaton AdS$_{3}$ 
gravity.

Finally,  one would like to extend our arguments to $d>3$ spacetime 
dimensions. Here the main question is whether or not  the interpolating 
feature of the charged BTZ black hole is a
peculiarity of $d=3$ and whether we can  extend it to a wide class of charged
and/or rotating black holes in $d>3$.

\end{document}